%% file: main.tex
\documentclass[conference]{IEEEtran}
\IEEEoverridecommandlockouts
\usepackage{cite}
\usepackage{amsmath,amssymb,amsfonts}
\usepackage{algorithmic}
\usepackage{graphicx}
\usepackage{textcomp}
\usepackage{xcolor}
\usepackage{rotating}
\usepackage{tikz}
\usepackage{enumitem}
\usepackage{url}
\usepackage{tcolorbox}
\usepackage{fontawesome}
\usepackage{balance}

\def\BibTeX{{\rm B\kern-.05em{\sc i\kern-.025em b}\kern-.08em
    T\kern-.1667em\lower.7ex\hbox{E}\kern-.125emX}}

\newcommand{\unitcode}{\textit{SIT723}}
\newcommand{\Ontrack}{\textit{OnTrack}}

\begin{document}

\title{Persona-based Assessment of Software Engineering Student Research Projects: An Experience Report}

%\author{\IEEEauthorblockN{Anonymous author(s)}}

\author{
    \IEEEauthorblockN{Chetan Arora\IEEEauthorrefmark{1}\IEEEauthorrefmark{2}, Laura Tubino\IEEEauthorrefmark{1}, Andrew Cain\IEEEauthorrefmark{1}, Kevin Lee\IEEEauthorrefmark{1}, Vasudha Malhotra\IEEEauthorrefmark{1} 
   }    
    \IEEEauthorblockA{\IEEEauthorrefmark{1}Deakin University, Geelong, Australia}
    \IEEEauthorblockA{\IEEEauthorrefmark{2}Monash University, Victoria, Australia}
    Email: chetan.arora@monash.edu, \{laura.t, andrew.cain, kevin.lee, v.malhotra\}@deakin.edu.au
    
%\vspace{-2em}
}

\newcommand{\sectopic}[1]{\vspace{0.2em}\par\noindent{\textit{\bfseries #1}}}

\maketitle

\input{Files/abstract.tex}
\input{Files/introduction.tex}

\input{Files/background.tex}

\input{Files/unitEditions.tex}
\input{Files/lessons.tex}

\input{Files/related.tex}
\input{Files/conclusions.tex}

\pagebreak
\bibliographystyle{IEEEtran}

\bibliography{paper}
\balance

\end{document}

%% file: Files/abstract.tex
\begin{abstract}
Students enrolled in software engineering degrees are generally required to undertake a research project in their final year through which they demonstrate the ability to conduct research, communicate outcomes, and build in-depth expertise in an area. Assessment in these projects typically involves evaluating the product of their research via a thesis or a similar artifact. However, this misses a range of other factors that go into producing successful software engineers and researchers. Incorporating aspects such as process, attitudes, project complexity, and supervision support into the assessment can provide a more holistic evaluation of the performance likely to better align with the intended learning outcomes. In this paper, we present on our experience of adopting an innovative assessment approach to enhance learning outcomes and research performance in our software engineering research projects. Our approach adopted a task-oriented approach to portfolio assessment that incorporates student personas, frequent formative feedback, delayed summative grading, and standards-aligned outcomes-based assessment. We report upon our continuous improvement journey in adapting tasks and criteria to address the challenges of assessing student research projects. Our lessons learnt demonstrate the value of personas to guide the development of holistic rubrics, giving meaning to grades and focusing staff and student attention on attitudes and skills rather than a product only.

% Categorizing performance expectations through the use of grade based persona

% process
% attitude/disposition
% cheating
% equity
% easy projects

% conduct communicate 
% lead
% working with researchers

% \textbf{Problems.}
% 1. No two projects are the same-equitable \\
% 2. Capture research skills as well as the quality of the research outputs\\
% 3. Conflict of interest from the supervisor;\\
% 5. Grades are usually arbitrary, no meaning; \\
% 6. Higher grades students are potentially our future generation of PhD students;\\

\end{abstract}

% Title: Keeping Fun Alive: an Experience Report on Running Online Coding Camps

% Abstract: The outbreak of the COVID-19 pandemic prohibited radically the collocation and face-to-face interactions of participants in coding bootcamps and similar experiences, which are key characteristics that help participants to advance technical work in coding camps. Several specific issues are faced and need to be solved when running online coding camps, which can achieve the same level of positive outcomes for participants. In this paper, we report on our experience and insights gained from designing and running a fully remote coding camp that exposes high school students to Agile-based Software Engineering practices to enhance their ability to develop high-quality software. To design the online coding camp, we adapted the face-to-face version of the bootcamp to keep the “level of fun” of the face-to-face coding camp, i.e., adaptations aimed at increasing communication, engaging participants, and introducing fun items to reduce fatigue due to prolonged computer use, while preserving the technical curriculum that enables students to attain the learning goals originally planned. The comparison with the results of the face-to-face coding camp shows that the two editions had the same effectiveness in terms of quality of product and process. From our experience, we synthesize lessons learned, and we sketch some guidelines for educators.

\begin{IEEEkeywords}
Software Engineering Education, Research Projects, Portolio-based Assessment
\end{IEEEkeywords}

%% file: Files/introduction.tex
\section{Introduction}~\label{sec:introduction}

%background to the situation
 Providing students with research experiences improves critical thinking and analytical skills, develops their self-confidence, allows them to feel part of the software engineering community \cite{hunter2007becoming}, and strengthens their employability profile \cite{hatziapostolou2018authentic}. Therefore, Software Engineering (SE) degrees commonly include a research component which results in a thesis \cite{knauss2021constructive,malachowski2018institutionalizing}.
%
% identify problem
Including research projects in SE degrees, however, has certain challenges which need resolution. Providing students with an authentic and equitable experience is difficult due to the variability of projects and the quality of supervisors. Students also present high variability in skills and understanding, with some being able to contribute intellectually to a research project idea and design and others needing clear instructions every step of the way. In addition, SE research projects commonly involve a practical problem in need of a solution. Doing high-quality empirical work within the time frame of a typical coursework research project is difficult~\cite{chakraborty2021towards}, and therefore requires a balance between academic values and practical relevance~\cite{knauss2021constructive}. These factors are among those that make assessing students' outcomes of research projects such a complex endeavor~\cite{bischof2011top}. 

% argue why the problem is important and needs to be solved
The implications of research project design and assessment not accurately reflecting the skills and achievements of students in these research projects are many, including student disengagement, attrition, and complaints. In addition, in our institution, the grade a student achieves in these research projects is considered when awarding PhD scholarships, making it a high-stakes issue. Along similar lines, the assessment in SE research projects reflects the ability of students to apply their overall learning in the SE degree in an ideally self-managed project. This has implications for the employment prospects of SE graduates and the reputational outcomes for the institution.

%summarise the solution

To address these issues, we redesigned the assessment in SE research projects, wherein students work on a SE research topic with one or more supervisors, in a one-to-one supervision model.  We worked on shifting the assessment focus, from the final thesis to the research process, through the submission of small regular tasks resulting in a thesis and a project management report. Scaffolding the process also ensures every student gets formative feedback throughout the teaching period. This process-oriented model is enhanced by SE research student personas that highlight to students the expectations at different achievement levels, and to research project supervisors the scaffolding needed by students at different levels. This model aims to address the variability of projects, students, and supervisors, resulting in a more accurate evaluation of a student's skills and achievements. 

In this paper, we describe and reflect on how we applied a design thinking approach~\cite{kelley2001art,plattner2010bootcamp} to research project assessment design. Each iteration is described, emphasising the evaluation and lessons learned. 
%establish credibility
The main contributions of this paper are (i) using a design thinking approach to redesigning SE research projects' assessment and (ii) exemplar assessment rubric and persona model, and our experiences deploying this exemplar assessment rubric and model; and (iii) our reflections and lessons learnt for developing a formative assessment model for improving the learning outcomes and research quality in SE research projects.

\sectopic{Structure.} Section~\ref{sec:background} provides context on Software Engineering research projects and describes our experiences. Section~\ref{sec:unitVersions} describes our assessment redesign approach using design thinking. Sections~\ref{sec:unit_v1},~\ref{sec:unit_v2} and~\ref{sec:unit_v3} present the redesign of the assessment model over three iterations. Section~\ref{sec:lessons} discusses the lessons learnt and our reflections on the three iterations. Section~\ref{subsec:LiteratureReview} addresses related work. Finally, Section~\ref{sec:conclusions} presents conclusions and future work.

%% file: Files/background.tex
\section{Context}~\label{sec:background}
In this section, we provide context to our university's teaching and learning environment, SE research projects, and motivate the case for redevelopment of SE research projects based on the issues in the existing model. 

\subsection{Bachelor of Software Engineering }~\label{subsec:SEContext}
Bachelor of Software Engineering (BSE) students complete a four-year degree, which prepares them for a career leading software development projects. In the first year, they complete technical units in traditional computer science, programming, software engineering, and project management. We note that a course, e.g., Java Programming 101, is termed as \emph{unit} in our context. In their second year, this foundation is expanded with more specialized SE skills such as AI and full-stack development. To utilize and provide evidence of their learning, students complete a technical capstone (industry-problem-based development team project) program in year three, as described in~\cite{schneider2020adopting}. 
%~\cite{schneider2020adopting}.
%
To develop and demonstrate an ability to use their skills for advanced research and development projects and contribute to the wider SE community, in their final year, students complete a SE research project unit (\unitcode).  %We note that, prior to~\unitcode, BSE students study another unit on research training. The unit aims to help them develop the skills necessary to undertake a SE research project, and be familiar with the general research methodologies. This unit also covers the major areas of SE research, and the students are encouraged to explore the area of research that is of most interest to them. %We have found that this hybrid approach of learning about research whilst exploring an area and developing a proposal allows lots of flexibility for Software Engineering students.  

%Kevin Lee - Can you please add a paragraph on the grading system - P/C/D/HD in Background? Also, if possible can you add a subsection in the background on the existing structure of the unit before SIT723/724, i.e., non OnTrack unit and only assessment of thesis. I think you should already have a template from Jan Carlo from the old marking scheme.
%\vspace*{.5em}
\subsection{SE Research Project (\unitcode) Unit}~\label{subsec:traditional}

As mentioned above, BSE students complete~\unitcode~in their final year. In~\unitcode, students work with one (or more) academic supervisors to undertake an ideally self-driven research project. This project is generally in an area of interest to the student and focused on demonstrating their learning outcomes of focused technical areas of SE research. The unit aims to help them develop research skills necessary to complete a SE research project, and be familiar with SE research process. The unit runs over all semesters at our university, with $\approx$100 students enrolled in it each semester. The unit includes research training in the form of workshops (e.g., literature search and review, academic writing and research project management) for all students by the unit coordination team, while in parallel individual students work with the academic supervisors on their specific research projects.

\sectopic{Project Allocation.} The allocation of students to the research projects begins with compiling a handbook of all research projects on offer for BSE students. All staff members in the department of Information Technology (IT) at our university propose research projects (collated in the handbook) for~\unitcode~students. The projects in the handbook range from exploratory (blue-sky) research to scoped projects, industry-style R\&D projects, or research projects with industry partners. The topics of the projects are diverse, such as requirements engineering (RE), applied SE in blockchain, cybersecurity, Internet of Things (IoT), Artificial Intelligence (AI) and other domains.  Each project's description would describe the overall scope of the project and the background/interest required to carry out the project, e.g., a project on ``RE for AI systems'' would require prior experience in addition to RE in machine learning (ML) and an interest in RE for AI systems~\cite{ahmad2021s,Ahmad:2022}.

The students enrolled in~\unitcode~provide their top-$N$ preferences from the handbook. The students are subsequently matched with the research projects based on their preferences and their background in the research project topics. The allocation process also takes students' overall average grade into consideration to avoid `over-booking' specific projects. For example, if students $S_x$ and $S_y$ are interested in the same project and have matching background, the one with higher average grade will be allocated to the project, assuming there is another matching project available for the other student.

\sectopic{Assessment Process.} At our university, the students outcomes are graded on a five level scale - below 50\% Fail ($\mathcal{F}$), 50\%-59\% Pass ($\mathcal{P}$), 60\%-69\% Credit ($\mathcal{C}$), 70\%-79\% Distinction ($\mathcal{D}$), and above 80\% High-Distinction ($\mathcal{HD}$). The actual numerical mark is calculated based on the quality of the assessed artefact. In BSE units, this has traditionally meant 2-3 main assessment artefacts, and possible examination, which are all assessed with defined marking schemes. 

In the past, in~\unitcode, the assessment was isolated to an academic's (research project supervisor's) review of the final thesis submitted by a student, based on a coarse marking grid. The marking scheme for~\unitcode~contained criteria for different sections of the submitted thesis, e.g., {\it overview}, {\it related work}, {\it research questions}, {\it methodology}, {\it evaluation; findings}, {\it interpretation}, {\it contributions}, {\it future work}, {\it references} and the overall {\it thesis presentation}. The final mark for each criterion, based on the quality produced, was awarded on a scale of 1-10, and totaling to 100. The final grade was then calculated based on the scale mentioned above.  %We present the improvements on~\unitcode~assessment model in Sections~\ref{sec:unitVersions}, \ref{sec:unit_v1}, \ref{sec:unit_v2}, and \ref{sec:unit_v3}.

%The pass level for each criteria is the minimum required to demonstrate the ability to perform fully directed research. Credit is awarded for generally producing good work across all of the thesis criteria. Distinction and High Distinction is awarded for a thesis that identifies and answers significant research questions or for substantial research contributions.

\sectopic{Assessment Issues in~\unitcode.}
% Students completing a Software Engineering honours research thesis qualify for entry to PhD. The grade in this unit also has a direct impact on the likelihood of a student gaining a PhD scholarship. Students with a final grade greater than 80\% have a high chance of being awarded a PhD scholarship. Due to the close interaction between the student and supervisor/mentor, it is also common that students will end up being supervised for a PhD by the same supervisor. 
%this needs to now be motivation...
The assessment model discussed above had been used in~\unitcode~for a long time and until the end of year 2020. The influence of assessment in~\unitcode~is not limited to only a final grade of the students, there is a far-reaching impact of the grade. The final grade of~\unitcode~has a significant weighting in the PhD scholarship allocation process at our university, which is highly competitive. Therefore, a recurring phenomenon with~\unitcode~is supervisors tendency to inflate the grades~\cite{Green:10}. There were a number of other issues in the existing assessment model of~\unitcode. The model unintentionally rewarded straightforward projects, to the detriment of more complex or exploratory ones. There were several relevant factors that were not accounted for, e.g., the variety of other documentation required for completing a research project, such as, work plans, work diaries and presentations, and the student's learning experience and skill development, such as, the extent of support provided to students, student's initiative in the project, and reflections on their learning~\cite{hazzan2004reflection}. We were motivated for improving the overall quality of the~\unitcode~students and BSE graduates, and addressing issues related to~\unitcode~including the ones mentioned above.

%% file: Files/unitEditions.tex
\section{Redesigning Assessment for the Research Project Units using a Design Thinking Approach}~\label{sec:unitVersions}

\begin{figure}[!t]
    \centering
    \vspace*{-.5em}    \includegraphics[width=0.47\textwidth]{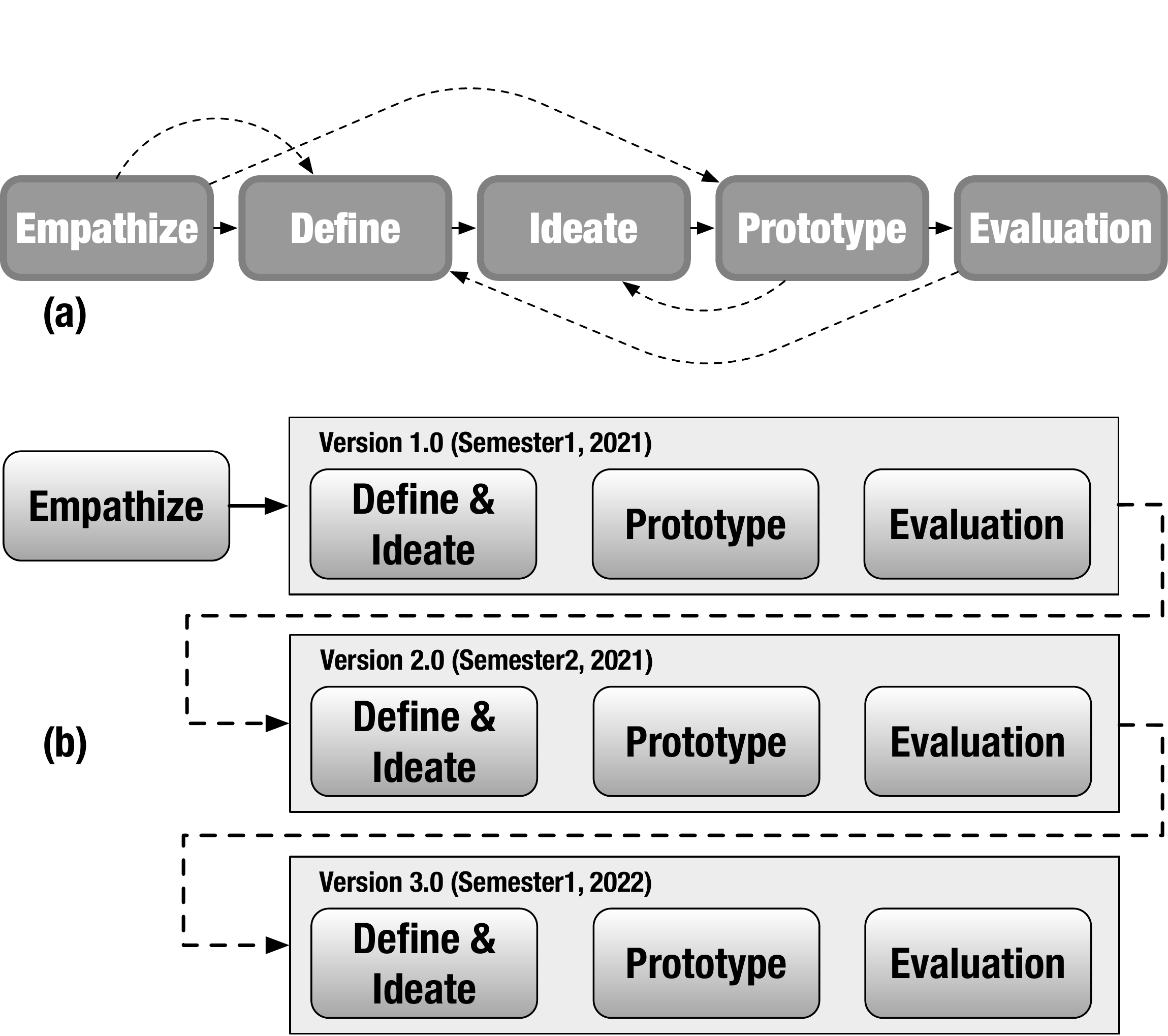}
    \vspace*{-.5em}
    \caption{(a) Design-thinking Approach~\cite{kelley2001art}; (b)~\unitcode~Redesign Process and Timeline based on the Design-thinking Approach.}\label{fig:DTProcess}
    \vspace*{-1em}
\end{figure}

In this section, we describe how we applied the design thinking approach in the evolution of the SE research project unit (\unitcode) at our university. We applied the five stages of the design thinking process~\cite{kelley2001art,plattner2010bootcamp} to improve the~\unitcode~design iteratively. Fig.~\ref{fig:DTProcess}~(a) shows an overview of the design thinking process. The first step \textbf{Empathize} focuses on understanding the key stakeholders and their concerns in the problem context. The next step is to \textbf{Define}, i.e., fully establish the stakeholders' concerns, synthesize these concerns and requirements, and correctly frame the problem at hand. The third step of \textbf{Ideate} focuses on brainstorming alternate solutions constructs, followed by the \textbf{Prototyping} of one of these alternate solutions. The last step is \textbf{Evalation}, where the prototype solutions are evaluated against the stakeholder requirements and concerns identified in the Empathize and the Design steps. We instantiated the steps iteratively (see Fig.~\ref{fig:DTProcess}~(b)) for redesigning~\unitcode. Below we report on all the steps in a sequence and the iterative improvement of the unit while redesigning, and the lessons learnt.

\subsection{Empathize Step}
In the first step (empathize), we focused on identifying the key issues with the stakeholders involved in~\unitcode~research projects. The key stakeholders are (i)~Bachelor of Software Engineering (BSE) students, (ii)~project supervisors, (iii)~the unit coordination team, and (iv)~the BSE management team. We identified the issues by reviewing the current unit content and structure (at the time), and having conversations with relevant stakeholders. There were numerous key issues (KIs) highlighted at this stage.

\sectopic{KI1 - Assessment focus.} The existing assessment model's analysis helped us identify the excessive focus on the end-product, i.e., ``research thesis'', in contrast to improving students' research skills. The assessment was conducted at the end of the semester, and the thesis was reviewed by the supervisors for a final grade. From a student's perspective, this gave them little opportunity to hone their research skills, with the research thesis being the main and the only assessed artefact. This also prevented students from getting intermediary feedback during their research projects. 

\sectopic{KI2 - Varied level of guidance required by students.} During our conversations with the supervisors, we determined that the level of support is a key variable factor in \unitcode. While some students are able to grasp the higher-level requirements of research projects and work independently, others expect a clearly laid out plan and are not able to have an abstract view of the research project as a whole. The overall unit structure, especially the assessment model, did not capture this factor to allow students to work at different levels of agency~\cite{biesta2007agency}. 

\sectopic{KI3 - Assessment equitability.} SE is a broad field of research with numerous sub-disciplines and applications. The diversity in the research projects and the lack of a common understanding of expected learning outcomes resulted in inequitable assessment. The issue was further exacerbated by the vested interest of supervisors in higher grades for students. For instance, in most editions of~\unitcode~prior to redesign, the fraction of students getting a $\mathcal{HD}$ was more than 60\%.
As noted in Section~\ref{sec:background}, the final mark in~\unitcode~had high influence on the allocation of otherwise highly-competitive PhD scholarships. This is a common issue in assessment fairness~\cite{Green:10}, and we identified clear patterns of `inflated grades' in the unit. From BSE management team's perspective, this was a reputational concern, as the students with high grades in~\unitcode~were being considered research ready and were entering SE PhD programme, when in reality they were not.

\section{\unitcode~Redesign~Version 1.0}~\label{sec:unit_v1}

Below, we present the first version (1.0)~\unitcode, redesigned based on the KIs identified above.

\subsection{Define and Ideate Steps}~\label{subsec:design_v1}
In the design step, we analyzed all the information collated from different stakeholders and the unit material, e.g., the unit assessment model, and grade distributions in the past. We further discussed ideas on ``how might we'' address the issues identified in the empathize step~\cite{o2020developing}. For the ideate part of the process, we discussed several alternative solutions to address each of the key issues established in the empathize step. Next, we discuss the results of our design and ideate steps.

\sectopic{Address KI1.}  In response to KI1, we discussed numerous research skills that we deemed necessary for SE research. %realizing that many of them can not be assessed by reading a student's thesis.
We had further discussions on structuring the unit assessment, and alleviating the issues of summative assessment at the end of the semester, based solely on the research thesis. It became clear that a more holistic approach to the management and assessment in~\unitcode~was required. 
%One alternative for addressing this was to utilize a portfolio-based approach~\cite{love2004designing,tubino2020authentic}. 
One proposition for addressing this issue was to use Task-Oriented Portfolio Assessment  (TOPA)~\cite{love2004designing,tubino2020authentic,cain2020using}. TOPA is a process-oriented assessment model. It addresses a few of our issues, such as the development of skills through its iterative feedback loops, holistic development through outcome-based assessment, and supporting the needs of our varied cohort through differentiated tasks. We discuss TOPA implementation in the next step.

\sectopic{Address KI2.} In response to KI2, we discussed introducing in our assessment model the concept of `agency'. We defined agency through a social cognitive lens as the ability to exert control over and direct ones own learning~\cite{biesta2007agency,etelapelto2013agency}. An important aspect of agency as viewed through a social cognitive perspective is its developmental nature~\cite{bandura2006adolescent}. Kegan's constructive developmental theory analyzes developmental changes through adulthood in terms of evolvement and exercise of human agency~\cite{kegan1994}. According to Kegan's stages of adult development, students in higher education present: instrumental, socialized and self-authoring minds (see Table~\ref{table:Kegan}). In our assessment design (discussed later in this section), we used Kegan's stages to understand how to better support and challenge our students holistically~\cite{kegan1994,magolda2004making}.

\sectopic{Addressing KI3.} In response to KI3, we wanted to introduce an assessment rubric, as part of the TOPA (discussed above). The assessment rubric should enable equitable assessment of students by focusing on research skills rather than research output only. Another aspect impacting the equitability of assessment was the issue of supervisor's control over their students' final grades, impacted by their vested interest in awarding high grades to secure PhD students and scholarships. To alleviate this issue, we discussed several alternatives, such as a SE conference program committee-style assessment process with no involvement of the supervisors, a moderation process involving a panel to moderate the final assessment conducted by supervisors, and a peer-review process.

\subsection{Prototype Step}~\label{subsec:prototypev1}
In this step, we implemented the~\unitcode~assessment based on student personas and TOPA. Below we present the student personas that guided our assessment model, the assessment rubric, TOPA design, and the grading process.

\input{Files/table_minds.tex}

\begin{figure}[!t]
    \centering
    %\vspace*{30em}
    %\hspace*{-70em}
    %\begin{tikzpicture}[scale=1]
        \includegraphics[width=\columnwidth]{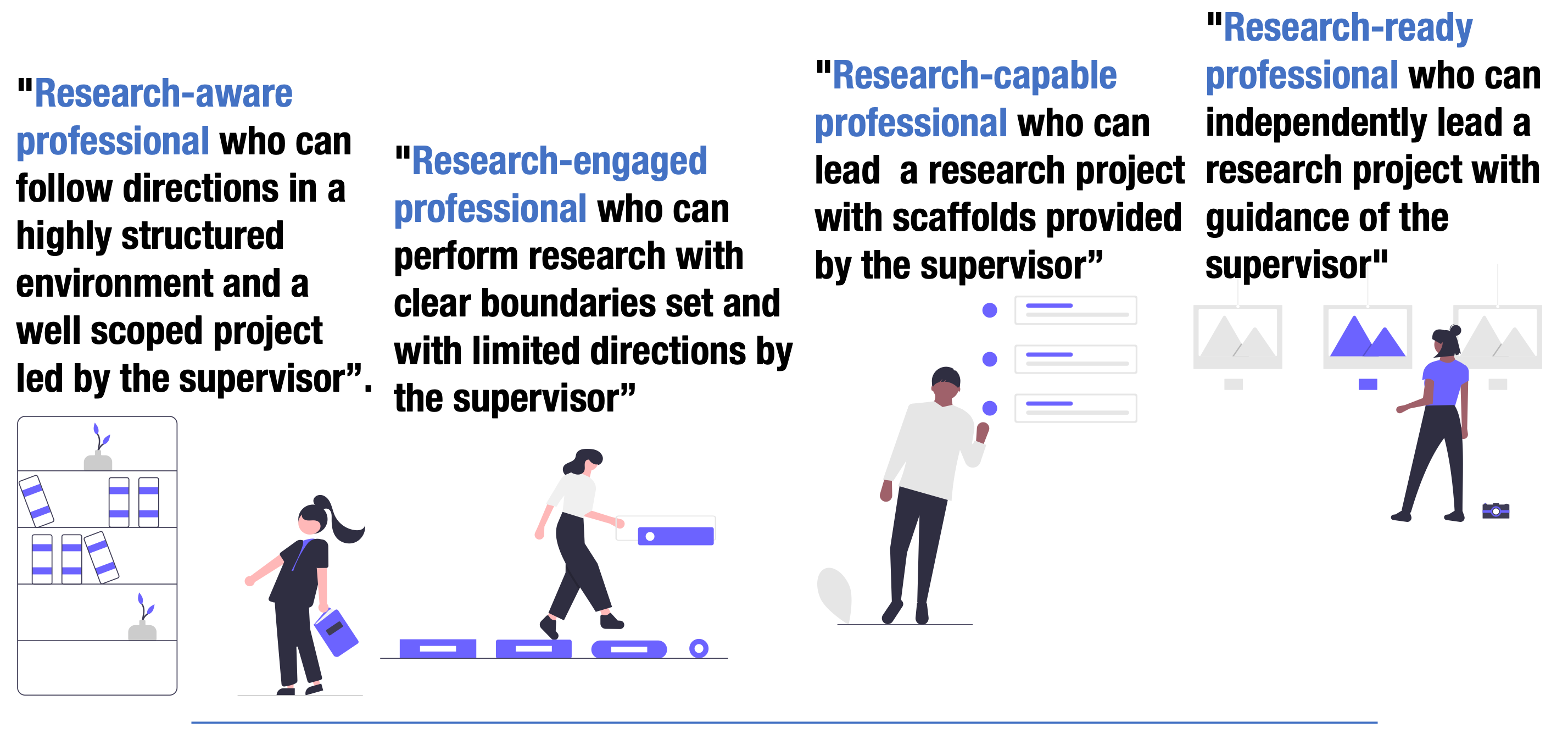}
    %\end{tikzpicture}
\caption{Research Personas of Students in \unitcode.} 
\label{fig:personas} 
\vspace*{-1em}
\end{figure}

\sectopic{Personas.} Based on the stages of adult development identified in Kegan's constructive-developmental theory \cite{kegan1994}  and present in higher education students \cite{magolda2004making} (Table~\ref{table:Kegan}), we created the ~\unitcode~personas (see Fig.~\ref{fig:personas}). These personas are used in the rubric design to guide us in identifying the different levels of achievement in a meaningful way, rather than arbitrarily. These personas are also used to convey expectations to both, students and supervisors.

$\bullet $ \emph{ Research-aware professional} describes a student at an instrumental mind stage. She requires an authority figure, in this case, a supervisor, to guide and scope the project for her. However, she is expected to be aware of all the stages of conducting research and should be able to communicate the project idea by the end of the semester. This student will achieve a Pass ($\mathcal{P}$) level according to our assessment design.

$\bullet $ \emph{Research-engaged professional} describes a student at a socialized mind stage. She relies on the supervisor for guidance and confirming the scope of the project. She has an understanding of all the stages of conducting research and is able to communicate the project idea. This student will achieve a Credit ($\mathcal{C}$) level in our assessment design.

$\bullet $ \emph{Research-capable professional} describes a student at a socialized mind stage. Once provided with an idea and clear direction, she can complete the project with limited involvement of the supervisor. She can conduct all the stages of research at an appropriate level and has a good understanding of the project idea from the start of the semester. This student will achieve a Distinction ($\mathcal{D}$) level.

$\bullet $ \emph{Research-ready professional} describes a student at a self-authoring mind stage. Alone, or in conjunction with the supervisor, she can define a research idea and a clear direction. She can complete the project with the supervisor as a `critical friend'. She can conduct all the stages of research and is able to communicate and discuss the project idea. This student will achieve a High-Distinction ($\mathcal{HD}$) level.

\begin{table*}[!h]
    \centering
    %\vspace*{30em}
    \hspace*{-5em}
    %\begin{tikzpicture}[scale=1]
    \caption{\unitcode~Assessment Rubric.}~\label{tab:Rubric} 
        \includegraphics[width=2.05\columnwidth]{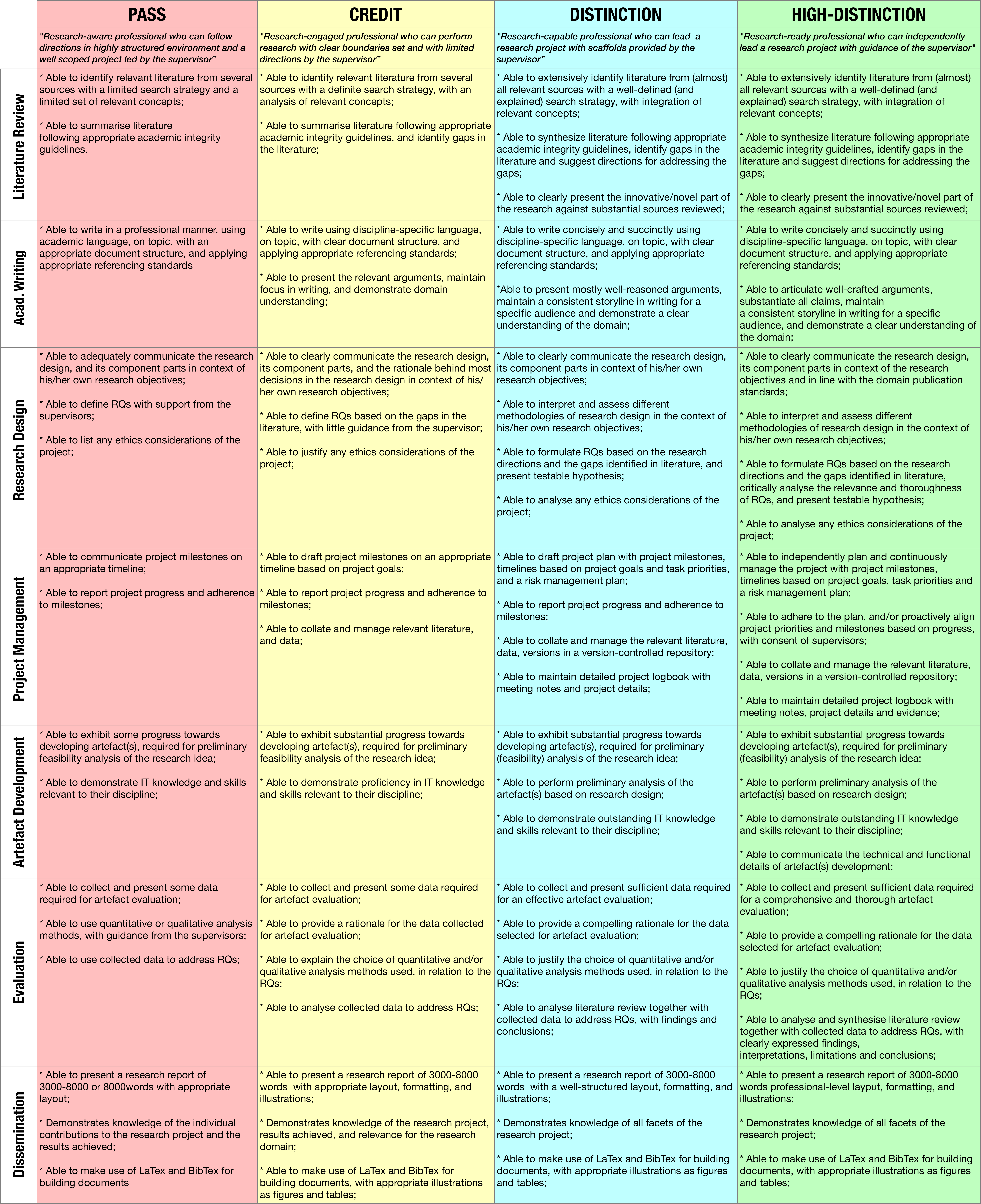}
    %\end{tikzpicture}
\end{table*}

\sectopic{Assessment Rubric.} Based on our discussions in the design and ideate steps, we first developed the assessment rubric aligned with the student personas and the research skills deemed relevant for~\unitcode. The assessment rubric and the criteria was developed based on the existing work on research skills development~\cite{willison2008researcher,willison2012academics,willison2016phd}, SE research and evaluation guidelines~\cite{Feldt:2009,kitchenham2002preliminary,Wohlin:12,stol2020guidelines}, Computer Science and SE teaching and learning guidelines~\cite{bavota2012teaching,Hazzan:15}.
Table~\ref{tab:Rubric} shows the assessment rubric we designed and implemented in \unitcode. 
The assessment rubric covers seven criteria (hereafter ARC\#1--7), namely, \emph{ARC\#1 - 
Literature Review, ARC\#2 - Academic (\& Technical) Writing, ARC\#3 - Research Design, ARC\#4 - Project Management, ARC\#5 - Artefact Development, ARC\#6 - Evaluation, and ARC\#7 - Research Dissemination}. The rubric presents descriptors for each criteria at all four grade levels. For a student to achieve a certain grade, they were required to match the grade for all criteria, i.e., to get a $\mathcal{HD}$, the student had to perform at the level across all criteria from the rubric.
%Overall, at a `Pass' level or a `Research Aware' student should be able to identify relevant literature (using a search strategy) based on the concepts relevant to the research direction. 
%The four personas of students, based on Kegan's developmental stages are: 

% \begin{figure*}[!t]
%     \centering
%     %\vspace*{30em}
%     %\hspace*{-70em}
%     %\begin{tikzpicture}[scale=1]
%         \includegraphics[width=1.8\columnwidth]{Figures/SIT723-724_TOPAMv1.png}
%     %\end{tikzpicture}
% \caption{TOPAM v1.} 
% \label{fig:TOPAMv1} 
% \end{figure*}

\begin{table}[!t]
    \centering 
    \caption{TOPA v1.0.}~\label{fig:TOPAMv1}  
        \vspace*{-1.5em}
        \includegraphics[width=0.85\columnwidth]{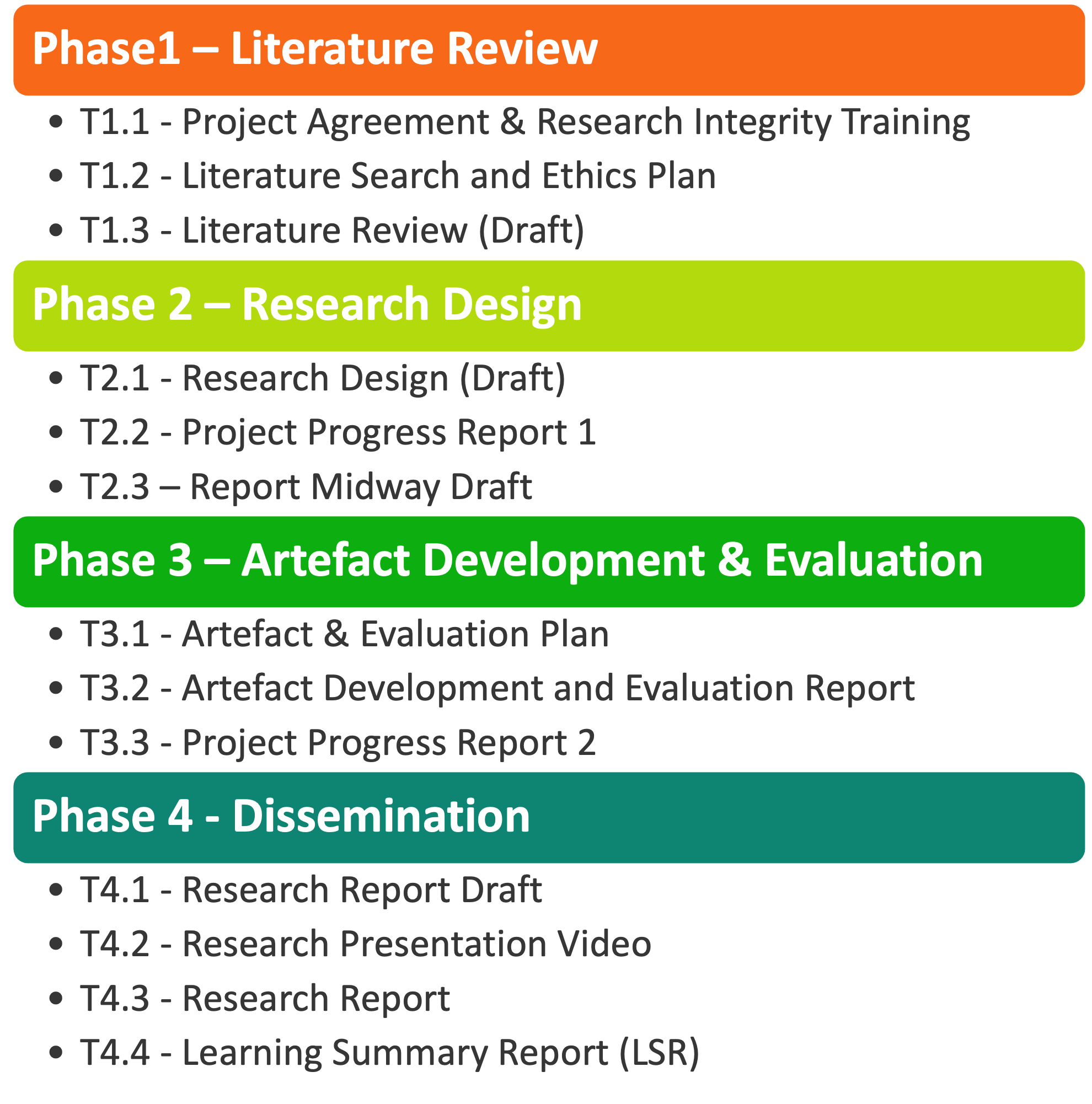}
    \vspace*{-1em}
\end{table}

\sectopic{Task-oriented Portfolio Assessment.} Based on the discussions in the empathize, design and ideate stages, we introduced TOPA\_v1 (see Table~\ref{fig:TOPAMv1}) where instead of students working towards a single block outcome, i.e., the research thesis, students would work on granular intermediate tasks throughout the semester -- which not only added up as their research portfolio, but also ensured all criteria in the rubric were taught, developed and assessed. As shown in Table~\ref{fig:TOPAMv1}, we split the project in four phases, discussed below. %We tried to map the assessment rubric criteria to our phases, with Project Management and Academic Writing being the criteria in all phases. \\

$\bullet $ \textbf{Phase 1: Literature Review} focused on project scoping and literature review, and contributed to criteria (ARC\#1, \#2, \#4, and \#7). In Phase~1, students had workshops to learn more about Research Processes, Literature Review and Project Management, and the workshops further offered them opportunities to discuss with their peers and the unit team, specifics of their own project.  The students were asked to submit following tasks on \Ontrack~-- our web-based outcome assessment system~\cite{cain2013examining,cain2020using}. All the tasks on~\Ontrack~were reviewed by the students' supervisors. \Ontrack~offers an option to mark a given task as complete, assign a grade for the graded tasks, or ask the student to resubmit the tasks after taking in the feedback provided for each task.
The tasks students worked on and submitted in this phase are:
\begin{itemize}
    \item \emph{T1.1 - Project Agreement and Training} - The students submitted a task on their project agreement with their supervisors, with the details of what needs to be done in the project based on an initial meeting between the student and the supervisor. The students were further required to complete a training module on research ethics and research integrity, and students uploaded the evidence of completion on~\Ontrack~in T1.1.
    \item \emph{T1.2 - Literature Search and Ethics Plan} - The students submitted their literature search strategy and results, e.g., the search terms and queries. The students were further asked to submit their ethics plan if their project required contact with any human participants. 
    \item \emph{T1.3 - Literature Review (Draft)} - The students were then asked to submit the draft of their literature review, for feedback from their supervisors at the end of Phase~1. 
\end{itemize}

$\bullet $ \textbf{Phase 2: Research Design} focused on developing the research design, i.e., planning the development of a relevant research artefact, formulating research questions (RQs), and planning the research evaluation. The students also addressed the supervisors' feedback on T1.3 from Phase~1. Phase~2 contributed primarily to criteria (ARC\#1, \#2, \#3, \#4, and \#7). In Phase~2, students had workshops on Formulating Research Questions and Artefacts and Evaluation in SE research. The tasks students worked on and submitted in this phase are:
\begin{itemize}
    \item \emph{T2.1 - Research Design (Draft)} - The students submitted the first draft of their RQs and the artefact development and evaluation plan to their supervisors on~\Ontrack.
    \item \emph{T2.2 - Project Progress Report~1} - The students submitted the overall progress report using a template provided to them. The students reflected on their progress, by positioning their progress to the assessment rubric criteria. As a part of the reflection process, the students self-assessed by grading themselves and providing evidence supporting their self-assessment. This was a graded task, so the supervisors assigned a grade and provided detailed feedback on students' reflections against the assessment rubric, helping them develop evaluative judgment skills~\cite{guest2022knowing}.
    \item \emph{T2.3 - Report Mid-Way Draft} - This task focused on a first draft of the research report formatted in \LaTeX. The students had access to an official template from the university on Overleaf\footnote{\url{https://www.overleaf.com/latex/templates/deakin-sit-thesis-template/tnqqzqtctmdw}}. The first draft was expected to include Introduction, Literature Review and Research Design, with supervisors' feedback from previous tasks addressed by the students. 
\end{itemize}

$\bullet $ \textbf{Phase 3: Artefact} focused on the development of a research artefact, e.g., a software prototype, algorithm, user interface design and survey design.   Phase~3 contributed primarily to criteria (ARC\#3, \#4, \#5 and \#7). In Phase~3, students had workshops on academic writing and research dissemination. The tasks students worked on and submitted in this phase are:

\begin{itemize}
    \item \emph{T3.1 - Artefact \& Evaluation Plan} - The students submitted (early on in Phase~3) a detailed plan for artefact development,  evidence of the required (programming) environment setup, and an artefact evaluation plan.
    \item \emph{T3.2 - Artefact Development and Evaluation Report} - The students submitted evidence of their artefact development completion and the artefact evaluation results. 
    \item \emph{T3.3 - Project Progress Report~2} - This task consisted of a progress report, similar to T2.2.
\end{itemize}

$\bullet $ \textbf{Phase 4: Dissemination} focused on the writing and dissemination phase, wherein the students completed their final research report. Phase~4 contributed to all criteria, as students revised all sections of their research report. No workshops were planned for this phase, except on-demand contact sessions with the unit team for clarifying doubts. The tasks students worked on and submitted in this phase are:

\begin{itemize}
    \item \emph{T4.1 - Research Report Draft} - The students submitted (early on in Phase~4) the completed draft of their research report in \LaTeX~to get feedback from their supervisors.
    \item \emph{T4.2 - Research Presentation Video} - In this task, the students recorded a three minute video describing their research project and their research contributions. The task was targeted at helping students convey their research outputs in a short time span. %The supervisors were instructed to assess the task as `Resubmit' on~\Ontrack, accompanied by actionable feedback if they were not satisfied with the submission.
    \item \emph{T4.3 - Research Report} - The students submitted their final research report after addressing the feedback from the supervisors in T4.1.
    \item \emph{T4.4 - Learning Summary Report (LSR)} - The students submitted their LSR, wherein they reflect on their overall learning and progress in \unitcode, including their target grade with an explanation of self-assessment against the assessment rubric~\cite{Tubino:20}. Once the LSR was submitted,~\Ontrack~enabled the students to automatically generate their `portfolio', that included all the previous task submissions and their supervisor's feedback.
\end{itemize}

\sectopic{Grading Process.} At the end of the semester, the student portfolios were assessed to assign a final grade.
From the three prospective solutions discussed for addressing KI3 in Section~\ref{subsec:design_v1}, we implemented an adaptation of peer-review process with supervisor as one of the participants.
Intermediary feedback to students on individual tasks in different phases (discussed above) during the teaching period was provided by the supervisor. The final grading on the portfolio was performed by three academics - the supervisor and two independent assessors. A large number of academics were involved in the process, as we attempted to engage as many people as possible. The three assessors for each portfolio did not know of each other's identity, and anonymously reviewed the portfolios. The distribution of the portfolios, random selection of independent assessors, and finalization of grades was managed by the unit team.
In case of a (major) discrepancy among the three academics, the unit team moderated the assessor's responses and assigned the final grade. 
%In case, the reviews from the two independent assessors agreed, then the grades were finalized using independent assessors' feedback and taking supervisor's comments on project management and student's overall persona into account. In case, one of the assessors agreed with the supervisors' review, that was deemed final, by taking any major criticism from the third assessor into account. In case all three assessors had divergent views, the unit team made the final decision.

\subsection{Evaluation Step}
We ran the redesigned~\unitcode~for the first time in the first semester of 2021. We collected verbal feedback from several supervisors on the unit design, both during the semester and at the end of the semester. Most supervisors appreciated the structure of the unit and the clarity provided by the personas and assessment rubric in guiding students. The~\unitcode~design definitely helped in addressing issues across KIs1--3. Anecdotally, in KI1, the focus on different criteria helped students focus on different aspects. For instance, most supervisors reported improved project management by students. For KI2, the personas and assessment rubric captured the varied level of guidance and helped aligning the student expectations, and several students self-assessed to adjust their target grades. For addressing KI3, we did receive positive feedback overall, but we were not sure of the extent to which the issue had been addressed.
This was due to the fact that the unit ran during COVID-19 lockdowns (in an online mode only), and only for a single iteration. The $\mathcal{HD}$ rate was $\approx$35\% in this iteration.
Having said that, there were a few further key issues (KIs) raised by several supervisors and the unit team.

\sectopic{KI4 - Flexibility in TOPA\_v1.} While the TOPA\_v1 task structure helped organize the intermediate tasks for students, students and supervisors felt restricted. The research process and progress in individual projects was not always aligned with the sequence of tasks in the TOPA\_v1 design. For example, some exploratory projects focused longer on literature review and lasted much longer across first three phases of Fig.~\ref{fig:TOPAMv1}.
\vspace*{.2em}
\begin{tcolorbox}[arc=0mm,width=\columnwidth,
                  top=1mm,left=1mm,  right=1mm, bottom=1mm,
                  boxrule=1pt] 
\faLightbulbO~\textbf{Flexibility in milestone sequencing} - Each SE research project is unique. The assessment model should focus on skills and intermediate milestones rather than the sequence or time-frame in which the milestones need to be achieved.
\end{tcolorbox}
\vspace*{.2em}

\sectopic{KI5 - Flexibility in the assessment rubric.}  The supervisors reported on the difficulty for students to perform well across all criteria, given the relatively short time span, the issues with the diversity of SE research projects in general, and project requirements. %For instance, some students performed surveys in~\unitcode~to establish the feasibility of the research problem. 
For example, a student working on the topic of ``practitioners perspective on AI-based software for regulatory compliance'', performed an extensive literature review and developed and conducted a survey with practitioners as her research project. While the project did not develop an artefact, typically considered in a SE project, e.g., a software prototype, the project had clear scientific contributions. Therefore, the assessment rubric required flexibility in its application. 

\vspace*{.2em}
\begin{tcolorbox}[arc=0mm,width=\columnwidth,
                  top=1mm,left=1mm,  right=1mm, bottom=1mm,
                  boxrule=1pt] 
\faLightbulbO~\textbf{Flexibility in the assessment rubric} - The flexibility of SE research projects and the limited time span of undergraduate research projects should be reflected in the assessment rubric. This is encouraged by the holistic assessment of the student's performance against the persona. %enable achieving higher grades for students even if not all criteria are at a higher level.
\end{tcolorbox}
\vspace*{.2em}

\section{\unitcode~Redesign~Version 2.0}~\label{sec:unit_v2}
\subsection{Design and Ideate Steps}~\label{subsec:design_v2}
After the first run of the unit in early 2021, we worked on addressing KI4 and KI5 from version~1.0. 
\sectopic{Address KI4.} While most supervisors and the unit team observed the benefits of the task-oriented model with intermediary feedback, the issues related to the sequencing of the milestones and flexibility in the assessment model required further changes. After rounds of discussions, we proposed a project-reporting based assessment model (TOPA\_v2), which used the TOPA\_v1 design in the background while offering flexibility to the supervisors and students. TOPA\_v2 required regular reporting of project progress, but the students and supervisors were free to sequence the milestones as they wished, as long as the final milestones aligned with the assessment rubric. The students were also provided information from TOPA\_v1 tasks as guiding material.

\sectopic{Address KI5.} Similar to TOPA\_v1, all stakeholders saw value in the assessment rubric, however some adjustments were required to offer flexibility to students to cater to the diverse SE research projects. We did not change the assessment rubric much, except a single change across all four grade levels for the Dissemination criterion. We made the use of~\LaTeX~as optional in the assessment rubric, as some students and supervisors were not comfortable with report writing in~\LaTeX. 

\subsection{Prototyping Step}
In this version of~\unitcode, we introduced the changes to the TOPA model (TOPA\_v2), and the rubric's implementation in assessment. The personas and the grading process remained unchanged. Below we present the changes.

% \begin{figure*}[!h]
%     \centering
%     %\vspace*{30em}
%     %\hspace*{-70em}
%     %\begin{tikzpicture}[scale=1]
%         \includegraphics[width=2\columnwidth]{Figures/SIT723-724_TOPAMv2.png}
%     %\end{tikzpicture}
% \caption{TOPAM v2.} 
% \label{fig:TOPAMv2} 
% \end{figure*}

\sectopic{Task-oriented Portfolio Assessment (TOPA\_v2).} Unlike, TOPA\_v1 that focused on sequencing intermediate milestones and tasks for SE research projects, TOPA\_v2 focused on project reporting on a weekly basis. Students submitted a weekly status reports, focused on reporting their weekly progress based on supervision meetings, summary of progress against the overall milestones, issues faced, and assistance required, e.g., issues related to accessing the university IT infrastructure. Students were encouraged to follow the four phases in TOPA\_v2, however, the tasks' submission did not enforce the phases. 
We retained TOPA\_v1 tasks T1.1, T2.2, T3.3, and T4.1-4.4 in TOPA\_v2.
% The structure of the weekly status report is: 
% \begin{enumerate}
%     \item Project Repository Link - link to all the project related work, including the literature, intermediary reports and artefacts;
%     \item Project Worklog - detailed worklog of the hours spent on the project;
%     \item Summary of the work planned for the previous week;
%     \item Summary of the work completed in the previous week;
%     \item Summary of any issues related to the progress and the assistance required, e.g., issues related to accessing the IT infrastructure of the university;
%     \item Summary of the work planned for the current week;
%     \item Overall milestones achieved and planned in the project;
% \end{enumerate}
%The project progress reports (T2.2 and T3.3 in TOPAM\_v1) were retained in TOPAM\_v2. However, in TOPAM\_v2, the students submitted a project progress report early-on in the project, which was deliberately targeted at ensuring that student had reviewed the assessment rubric, and were aware of expectations at different grade levels. 
The project progress report tasks (T2.2 and T3.3) were formatively graded, i.e., students received an indicative grade from their supervisors along with the feedback on their progress against the assessment rubric.

\sectopic{Assessment Rubric Implementation.} As discussed in Section~\ref{subsec:design_v2}, we did not change much in the assessment rubric criteria, however, we did change the implementation of the rubric and the related expectations. While some supervisors assessed based on the persona profiles, most took the rubric literally. Based on supervisors' feedback (KI5), we lowered the expectation of achieving a matching grade across all criteria to be eligible for that overall grade. This was particularly relevant for the Academic Writing criterion, as we have hundreds of international students from different nationalities, with English as their second or the third language, who were struggling to perform at higher levels in academic writing.

We marked Project Management and Dissemination (ARC\#4 and ARC\#7) as core criteria for which the students had to match the grade, e.g., for an overall $\mathcal{D}$, they had to get at least the Distinction in Project Management and Dissemination. For the remaining five criteria (ARC\#1, ARC\#2, ARC\#3, ARC\#5 and ARC\#6), we had following conditions for achieving different overall grade levels:
\begin{itemize}
    \item \textbf{Pass.} All five criteria at $\mathcal{P}$;
    \item \textbf{Credit.} At least three of these five criteria at $\mathcal{C}$. The remaining two criteria at $\mathcal{P}$ or higher; 
    \item \textbf{Distinction.} At least two of the five criteria at $\mathcal{D}$. The remaining two criteria at $\mathcal{C}$ or higher;
    \item \textbf{High Distinction.} At least one of these criteria at $\mathcal{HD}$. From the remaining criteria, at least two need to be at $\mathcal{D}$ or higher. All remaining criteria need to be at $\mathcal{C}$ or higher.
\end{itemize}

If a student ($S_x$) had the following grades, 
ARC\#1~($\mathcal{C}$),
ARC\#2~($\mathcal{C}$),
ARC\#3~($\mathcal{D}$),
ARC\#4~($\mathcal{HD}$),
ARC\#5~($\mathcal{HD}$),
ARC\#6~($\mathcal{D}$), and
ARC\#7~($\mathcal{HD}$), they would get an overall $\mathcal{HD}$ based on the conditions mentioned above. Let's say another student ($S_y$) had the same grades as $S_x$ for ARCs\#1--5 and ARC\#7, except a $\mathcal{HD}$ in ARC\#6 (instead of $\mathcal{D}$), they would also get a $\mathcal{HD}$ grade but a higher mark (out of 100) than $S_x$. So, $S_x$ would receive a mark of 80 ($\mathcal{HD}$), and $S_y$ would get a mark of 85 ($\mathcal{HD}$).

\subsection{Evaluation Step}~\label{subsec:evaluation_v2}
We implemented the next version of~\unitcode~in the second half of 2021. We again collected the verbal feedback over the semester from the unit team and supervisors on the unit design and the new changes. All stakeholders reacted extremely positively to the changes introduced in TOPA\_v2 and the flexibility introduced in the implementation of the assessment rubric, i.e., the KI4 and KI5 had been addressed to a large extent. At the end of the semester, we further observed the grading process (KI3) and discovered that issues related to assessor bias~\cite{Green:10,yeates2013you} had not been mitigated. The situation was better than before the redesign (version 1.0), i.e., supervisors' solely grading the students but there were still remaining issues. 

\sectopic{KI6 - Administrative Issues and Bias in the Grading Process.}
There were numerous issues in the previous grading process. Firstly, from an administrative perspective managing the process described in Section~\ref{subsec:prototypev1} for hundreds of students in a semester was taxing and cumbersome for the unit team. Secondly, the involvement of 40-50 assessors and anonmymization attempts in the grading process, further exacerbated the administrative issues. Finally, supervisor's bias still played a substantial role in the final grade as the supervisor was an active part of the grading process. Due to these issues, the overall implementation of the units redesign suffered as the unit team struggled to effectively moderate all the portfolios in this process, and the supervisors' bias was still influential.

\vspace*{.2em}
\begin{tcolorbox}[arc=0mm,width=\columnwidth,
                  top=1mm,left=1mm,  right=1mm, bottom=1mm,
                  boxrule=1pt] 
\faLightbulbO~\textbf{Getting the grading process right is difficult} - The grading process needs to be carefully designed, not only to be equitable but also to be feasible within the university's contextual constraints such as the staff workload issues. 
\end{tcolorbox}
\vspace*{.2em}

\section{\unitcode~Redesign~Version 3.0}~\label{sec:unit_v3}

\subsection{Design and Ideate Steps}~\label{subsec:design_v3}
In version~3.0, we primarily worked on addressing KI6. 

\sectopic{Address KI6.} We discussed several alternatives for improving the grading process and addressing the issues with supervisors' bias, within our organizational context and constraints, e.g., the workload available for the staff members for the grading process and the duration of assessment for a given solution. Our alternatives included, among others, (i) continuous assessment by an external assessor, however, this was deemed infeasible due to the workload (scalability) issues with hundreds of students in~\unitcode~in a semester; and (ii) a grading system similar to a conference peer-review system~\cite{ernst2021understanding}. We implemented a customized version of the (ii)~alternative.

\subsection{Prototype Step}~\label{subsec:prototype_v3}
In this version of~\unitcode, we introduced the changes to the grading process. All other aspects of the unit, e.g., the TOPA\_v2, personas and the assessment rubric seemed to work in a stable mode -- without any major issues. The new grading moderation process works as follows:
\begin{itemize}
\item At the end of the semester, the supervisor assigns a tentative grade and provides detailed comments for each criterion based on the rubric in~\Ontrack. We note that the end of the semester assessment from the supervisor, is in addition to the weekly and milestones-based feedback. 
\item The student's research portfolio, self-assessment of the student against the assessment rubric in LSR, and the supervisor's comments against the assessment rubric are passed onto a ``moderation panel''. 
\item The panel consists of $\approx$10 staff members, and functions very similar to the conference program committees. The panel convenes over a working day, and each member receives workload for $\approx$1.5 working days for participation.
\item The primary task of panel members is to \emph{moderate} the supervisors' tentative grade based on detailed comments and student's self-assessment and the portfolio. The panel members do not \textit{mark} the portfolio to address workload issues, i.e., the assessment of hundreds of students by multiple staff members is not scalable. Each portfolio is first moderated by a pair of panel members. Thereafter, the grade is finalized based on:
\begin{itemize}
\item If the pair agrees with each other on the final grade and the mark is the same as  the supervisor's grade $\longrightarrow$ the grade is finalized with no further discussion;
\item If the pair agrees on the final grade, but the grade is not the same as supervisor's grade $\longrightarrow$ the case will be briefly discussed among all panel members, and the pair might be asked to justify the grade change;
\item If the pair disagrees with each other on the final grade, then a third panelist is asked to review the grade and reach consensus independently $\longrightarrow$ the case is also be discussed with the entire panel;
\end{itemize}
\end{itemize}

\subsection{Evaluation Step}~\label{subsec:evaluation_v3}
In the first semester of 2022, we implemented version 3.0. The grading process seemed to work much better than the previous version, as supervisor was not an active part of the process. The overall impact of the supervisors' influence had decreased, and the overall $\mathcal{HD}$ rate had significantly decreased as compared to the version before version~1.0. We note that, due to the limited availability of staff members and workload-related issues, we only ran the moderation process for portfolios assessed by supervisors as $\mathcal{F}$, $\mathcal{D}$ or $\mathcal{HD}$. For the portfolios graded by supervisors as $\mathcal{P}$ or $\mathcal{C}$, the supervisors' grade was final. The rationale behind this is that only students with $\mathcal{D}$ or $\mathcal{HD}$ are eligible for a PhD entrance, and $\mathcal{F}$ grades required further review to ensure the fairness in the process. While this version is not perfect, it is a significant improvement over the previous versions in addressing supervisor grading bias, according to the unit team and the moderation panel. 

At the end of version~3.0, there are still some issues, e.g., supervisors still being responsible for providing intermediate grades on project progress reports (T2.2 and T3.3 in Table~\ref{fig:TOPAMv1}) and the final portfolio, which need addressing and we will attempt to address them in the future semesters. 

%\vspace*{.2em}
\begin{tcolorbox}[arc=0mm,width=\columnwidth,
                  top=1mm,left=1mm,  right=1mm, bottom=1mm,
                  boxrule=1pt] 
\faLightbulbO~\textbf{Supervisor's should focus on providing feedback rather than grading} - Supervisors should be responsible for mentoring and providing critical feedback to students at each stage of the process and for facilitating the moderation process. Grading can take into account supervisors' views but they should not be an active part of the grading process. 
\end{tcolorbox}
%\vspace*{.2em}

%\vspace*{.2em}
\begin{tcolorbox}[arc=0mm,width=\columnwidth,
                  top=1mm,left=1mm,  right=1mm, bottom=1mm,
                  boxrule=1pt] 
\faLightbulbO~\textbf{Fidelity of process implementation is challenging due to variability of supervisors} - The fidelity of implementation of the assessment model remains a challenge with the variability of supervisors. %This is likely due to the disengagement of a fraction of supervisors. 
Due to this, it is difficult to maintain  assessment consistency and equitability. 
\end{tcolorbox}
\vspace*{.2em}

%% file: Files/table_minds.tex
\begin{table}[!t]
%\normalsize
\centering
\caption{Robert Kegan's developmental stages}
\label{table:Kegan}
\vspace*{-1em}
\begin{center}
\begin{tabular}{ |p{1.8cm}|p{6.2cm}| }
 \hline
 \textbf{Developmental stage} & \textbf{Characteristics}
 \\
 \hline
 \textbf{Instrumental mind}  
 & Students commonly make decisions based on the benefits or consequences that may result from these, thus needing authority figures or rules to direct them.  
 \\
  \hline
\textbf{Socialized mind}  
 & Students are ``good citizens'', taking others into account and considering long term consequences. They understand others' points of view, even if different from their own.  
 \\
\hline
\textbf{Self-authoring mind}  
 & Students can evaluate arguments (their own and others’) using evidence, take contextual factors into account, acknowledge multiple ways of framing arguments, and reevaluate the basis of a decision as evidence changes.  
\\
  \hline
 \end{tabular}
 \end{center}
 \vspace*{-1em}
 \end{table}

%% file: Files/lessons.tex
\vspace*{-.5em}
\section{Lessons Learnt}~\label{sec:lessons}
\vspace*{-.5em}

We started this project to improve the reliability of our assessment of final-year SE research projects, which demonstrated challenges in terms of variability in students' skills, nature of projects, and quality of supervision. Through the process of developing and refining the assessment model for~\unitcode~we have largely been able to develop an assessment model that largely addresses the challenges associated with students and projects, and we are working on helping support staff to ensure the quality of supervision improves.

Personas are helping clarify what is expected of students, with grade-level descriptors assisting in communicating the standard of work required from the student and the support required from the supervisor. Rubric descriptors encourage students to seek greater agency and take more responsibility for their learning and research project. This helps demonstrate the benefits of the development rubric, which links assessment results with desired outcomes. Example outcomes of \unitcode~for a student with high levels of agency are~\cite{olorunnife2021automatic,wang2021virtual,lefevre2022modelops,CobbQuantum2022}, which were accepted for publication in peer-reviewed venues.

The flexibility provided by TOPA and enacted by adjusting the task requirements in v2.0 has allowed the model to cater to a wide variety of project types associated with SE research, from highly practical to fully theoretical. In assessing the research performance holistically, the challenge is to provide a sufficient structure within the assessment to have defined feedback points without overly constraining the kinds of projects that students can engage with. By shifting to a flexible task structure, we have been able to support students with formative feedback while allowing a wide variety of projects and an overall assessment that takes into consideration the wide range of factors that make up successful SE research projects. Furthermore, the TOPA\_v2 has been great in providing space for academics to make adjustments to assessment without requiring adjustments to published curriculum documentation.

While the assessment rubric has been useful in capturing and representing criteria in a traditional two-dimensional table structure, it requires adjustments to encourage the holistic assessment aimed for. Reactively adjusting the criteria needed to achieve specific grades helped to mitigate some of the issues we experienced. However, this highlights the need for a rethink of the presentation of the rubric, which may assist in achieving greater alignment with intended grade-level outcomes.

Variability in the quality of supervision is a remaining challenge. Implementing this model at scale has involved working with a large number of supervisors, and highlighted the difficulties of designing assessment to be delivered by academics who are not involved in the design \cite{tubino2021reforming}. Moderation processes can aid in identifying and addressing some of the issues that arise from this, mitigating its impact on final results. However, we need to rethink aspects of the model to ensure that we are able to identify and support students earlier in the process - thereby ensuring they have the support they require.

More generally, this continuous improvement process has demonstrated the importance of flexibility in the assessment model. Assessment within the unit has retained the same assessment model across all iterations, consisting of frequent formative feedback, delayed summative grading, and standards-aligned outcomes-based assessment. By making iterative changes to the assessment strategy, we have been able to improve the fidelity of the implementation, progressively adjusting the unit to address challenges that have arisen.

%% file: Files/related.tex
\vspace*{-.1em}
\section{Related Work}~\label{subsec:LiteratureReview}
\vspace*{-.5em}
%In this section we review the existing body of work on research projects in SE education, including any research strands on Master's and Bachelor's thesis and research training of SE graduates, and the existing work on task-oriented portfolio assessment (TOPAM) model.

%\subsection{Student Research in SE Education}
A large body of work exists on students research projects in general~\cite{healey2009developing,matzen2012defining,linn2015undergraduate,mcmanus2019project}, students group-based SE capstone projects~\cite{paasivaara2018does,paasivaara2019collaborating,tubino2020authentic}, however only a few research strands have covered student research projects design or research training in SE. Bernat et al.~\cite{bernat2000structuring} present a model on SE student research training and engagement, with the close collaboration of SE research groups and the students. In this model, each student define activities and timelines for a task assigned to them. The assigned tasks are deliverables (artefacts), such as literature review, software products or documentation. Similar to our assessment rubric, this model also focuses on helping students develop dissemination and writing skills. 

On similar lines, the structure of a SE Master Thesis program has been discussed in detail for a Swedish university~\cite{dodig2009professional,Feldt:2009,dodig2010improved,Host:2010}. In these papers, the authors present the idea of using rubric-based assessment, wherein four rubrics cover the research proposal, research project process, research thesis, and oral presentation. Similar to our assessment rubric of Table~\ref{tab:Rubric}, their rubrics state quality criteria and
different levels (superior, good, fair and minimal) of quality for each criterion. While the authors mention these rubrics in their papers, to the best of our knowledge, the complete versions of the rubrics are not available publicly. A subset of the rubrics on timeliness (project), balance (project), and ethical issues (thesis) is presented~\cite{dodig2009professional}. 
Seyed-Abbassi~\cite{seyed2006practical} reports on the implementation of SE research projects in a database unit, wherein the students worked in groups on literature review, software artefacts, including software requirements, implementation, academic writing and dissemination via a report, poster demonstrations and peer presentations. This paper also presents the assessment of the research project by assigning points to each intermediate task, and the points are aggregated at the end of the semester by the unit team. 

Similar to the papers discussed above, a few strands exist on (i) collaboration with the research groups and industry for improving students' research skills~\cite{shaw1998research,bargh2014research}, and (ii) research training and skills on specific SE tasks, such as automated code generation~\cite{khmelevsky2012automatic}, and software testing~\cite{oyeniran2021environment}, but none of these papers present details on the actual design of the unit and the assessment, which is the Achilles's heel of implementing a research unit or program, like~\unitcode.

% \subsection{TOPAM}
% \cite{renzella2017supporting} task-oriented portfolio
% assessment in supporting introductory programming units

% Value in reflection in SE teaching and learning~\cite{hazzan2004reflection}

% - Group Projects \\
% - ICSE 2021 paper - ``Reforming assessment: challenges beyond design'' 

%% file: Files/conclusions.tex
\vspace*{-.2em}
\section{Conclusions}~\label{sec:conclusions}
\vspace*{-.2em}
In this paper, we reflect on the evolution of our assessment model for SE research project units through a design thinking approach. The main challenges this model aims to address consist of catering to a wide range of projects, students' skills, and supervisor quality.
Key lessons indicate the importance of designing a flexibile and holistic approach for assessing research projects. Our focus on student agency (personas) and on having continuous, process-focused tasks that allow students to report on different areas at regular intervals, provided both the flexibility and structure required by the range of research projects present in SE. Due to the impact that students' grades in this units have on their opportunities to pursue a PhD, the grading process became a strong focus of this redesign. Our moderation processes were able to address the challenges of providing reliable grades, however, we still need to find a timely way to identify and address the challenges of differences in quality of supervision. This is essential for ensuring equitability of opportunity in such a high stake unit.

Going forward we will continue using this design thinking approach. In the next iteration we will go through the empathize stage, this time focusing on empathizing with the supervisors' experience of the assessment.

% \begin{itemize}
%     \item Students work is R\&D;
%     \item Task-oriented approach for providing intermediate feedback;
%     \item What is the assessment;
%     \item What was the problem - 
%     \item conflict of interest with the students on marks;
%     \item Brief literature on - 
%     \item Adding moderation - 
%     \item Big picture for the assessment rubric; 
%     \item Continuous assessment helps assess the process;
%     \item Learning outcomes - 
%     \item I did this vs my supervisor did the other things - contributions of the supervisor vs student's contributions; 
%     \item The persona only sets the potential, not the result;
%     \item Hollistic assessment - research output, attitude,  process, skills;
%     \item Persona and Product in the rubric;
%     \item Works well in SE - artefact and the thesis;
%     \item 
% \end{itemize}

% \subsection{Future Work}
% \begin{itemize}
%     \item Intermediary Panel Discussions for HD students;
% \end{itemize}

%% file: main.bbl
% Generated by IEEEtran.bst, version: 1.14 (2015/08/26)
\begin{thebibliography}{10}
\providecommand{\url}[1]{#1}
\csname url@samestyle\endcsname
\providecommand{\newblock}{\relax}
\providecommand{\bibinfo}[2]{#2}
\providecommand{\BIBentrySTDinterwordspacing}{\spaceskip=0pt\relax}
\providecommand{\BIBentryALTinterwordstretchfactor}{4}
\providecommand{\BIBentryALTinterwordspacing}{\spaceskip=\fontdimen2\font plus
\BIBentryALTinterwordstretchfactor\fontdimen3\font minus
  \fontdimen4\font\relax}
\providecommand{\BIBforeignlanguage}[2]{{%
\expandafter\ifx\csname l@#1\endcsname\relax
\typeout{** WARNING: IEEEtran.bst: No hyphenation pattern has been}%
\typeout{** loaded for the language `#1'. Using the pattern for}%
\typeout{** the default language instead.}%
\else
\language=\csname l@#1\endcsname
\fi
#2}}
\providecommand{\BIBdecl}{\relax}
\BIBdecl

\bibitem{hunter2007becoming}
A.-B. Hunter, S.~L. Laursen, and E.~Seymour, ``Becoming a scientist: The role
  of undergraduate research in students' cognitive, personal, and professional
  development,'' \emph{Science education}, vol.~91, no.~1, pp. 36--74, 2007.

\bibitem{hatziapostolou2018authentic}
T.~Hatziapostolou, D.~Dranidis, A.~Sotiriadou, P.~Kefalas, and
  I.~Nikolakopoulos, ``An authentic student research experience: fostering
  research skills and boosting the employability profile of students,'' in
  \emph{Proceedings of the 23rd Annual ACM Conference on Innovation and
  Technology in Computer Science Education}, 2018, pp. 254--259.

\bibitem{knauss2021constructive}
E.~Knauss, ``Constructive master's thesis work in industry: guidelines for
  applying design science research,'' in \emph{2021 IEEE/ACM 43rd International
  Conference on Software Engineering: Software Engineering Education and
  Training (ICSE-SEET)}.\hskip 1em plus 0.5em minus 0.4em\relax IEEE, 2021, pp.
  110--121.

\bibitem{malachowski2018institutionalizing}
M.~Malachowski, J.~M. Osborn, K.~K. Karukstis, J.~Kinzie, and E.~L. Ambos,
  ``Institutionalizing undergraduate research and scaffolding undergraduate
  research experiences in the stem curriculum,'' in \emph{Best Practices for
  Supporting and Expanding Undergraduate Research in Chemistry}.\hskip 1em plus
  0.5em minus 0.4em\relax ACS Publications, 2018, pp. 259--269.

\bibitem{chakraborty2021towards}
S.~Chakraborty, L.~Deng, and J.~Dehlinger, ``Towards authentic undergraduate
  research experiences in software engineering and machine learning,'' in
  \emph{Proceedings of the 3rd International Workshop on Education through
  Advanced Software Engineering and Artificial Intelligence}, 2021, pp. 54--57.

\bibitem{bischof2011top}
H.-P. Bischof, J.~D. Furst, D.~S. Raicu, and S.~D. Ruban, ``Top issues in
  providing successful undergraduate research experiences,'' in
  \emph{Proceedings of the 42nd ACM technical symposium on Computer science
  education}, 2011, pp. 239--240.

\bibitem{kelley2001art}
T.~A. KELLEY, \emph{The art of innovation: Lessons in creativity from IDEO,
  America's leading design firm}.\hskip 1em plus 0.5em minus 0.4em\relax
  Broadway Business, 2001, vol.~10.

\bibitem{plattner2010bootcamp}
H.~Plattner, ``Bootcamp bootleg,'' \emph{Design School Stanford, Palo Alto},
  2010.

\bibitem{schneider2020adopting}
J.-G. Schneider, P.~W. Eklund, K.~Lee, F.~Chen, A.~Cain, and M.~Abdelrazek,
  ``Adopting industry agile practices in large-scale capstone education,'' in
  \emph{2020 IEEE/ACM 42nd International Conference on Software Engineering:
  Software Engineering Education and Training (ICSE-SEET)}.\hskip 1em plus
  0.5em minus 0.4em\relax IEEE, 2020, pp. 119--129.

\bibitem{ahmad2021s}
K.~Ahmad, M.~Bano, M.~Abdelrazek, C.~Arora, and J.~Grundy, ``What’s up with
  requirements engineering for artificial intelligence systems?'' in \emph{2021
  IEEE 29th International Requirements Engineering Conference (RE)}.\hskip 1em
  plus 0.5em minus 0.4em\relax IEEE, 2021, pp. 1--12.

\bibitem{Ahmad:2022}
\BIBentryALTinterwordspacing
K.~Ahmad, M.~Abdelrazek, C.~Arora, M.~Bano, and J.~Grundy, ``Requirements
  engineering for artificial intelligence systems: A systematic mapping
  study,'' 2022. [Online]. Available: \url{https://arxiv.org/abs/2212.10693}
\BIBentrySTDinterwordspacing

\bibitem{Green:10}
S.~K. Green and R.~L. Johnson, \emph{Assessment is essential}.\hskip 1em plus
  0.5em minus 0.4em\relax Mcgraw-Hill Higher Education, 2010.

\bibitem{hazzan2004reflection}
O.~Hazzan and J.~E. Tomayko, ``Reflection processes in the teaching and
  learning of human aspects of software engineering,'' in \emph{17th Conference
  on Software Engineering Education and Training, 2004. Proceedings.}\hskip 1em
  plus 0.5em minus 0.4em\relax IEEE, 2004, pp. 32--38.

\bibitem{biesta2007agency}
G.~Biesta and M.~Tedder, ``Agency and learning in the lifecourse: Towards an
  ecological perspective,'' \emph{Studies in the Education of Adults}, vol.~39,
  no.~2, pp. 132--149, 2007.

\bibitem{o2020developing}
G.~O'Callaghan and C.~Connolly, ``Developing creativity in computer science
  initial teacher education through design thinking,'' in \emph{United Kingdom
  \& Ireland Computing Education Research conference.}, 2020, pp. 45--50.

\bibitem{love2004designing}
T.~Love and T.~Cooper, ``Designing online information systems for
  portfolio-based assessment: Design criteria and heuristics,'' \emph{Journal
  of Information Technology Education: Research}, vol.~3, no.~1, pp. 65--81,
  2004.

\bibitem{tubino2020authentic}
L.~Tubino, A.~Cain, J.-G. Schneider, D.~Thiruvady, and N.~Fernando, ``Authentic
  individual assessment for team-based software engineering projects,'' in
  \emph{2020 IEEE/ACM 42nd International Conference on Software Engineering:
  Software Engineering Education and Training (ICSE-SEET)}.\hskip 1em plus
  0.5em minus 0.4em\relax IEEE, 2020, pp. 71--81.

\bibitem{cain2020using}
A.~Cain, L.~Tubino, and S.~Krishnan, ``Using technology to enable a shift from
  marks to outcomes-based assessment,'' in \emph{Re-imagining University
  Assessment in a Digital World}.\hskip 1em plus 0.5em minus 0.4em\relax
  Springer, 2020, pp. 229--245.

\bibitem{etelapelto2013agency}
A.~Etel{\"a}pelto, K.~V{\"a}h{\"a}santanen, P.~H{\"o}kk{\"a}, and S.~Paloniemi,
  ``What is agency? conceptualizing professional agency at work,''
  \emph{Educational research review}, vol.~10, pp. 45--65, 2013.

\bibitem{bandura2006adolescent}
A.~Bandura, ``Adolescent development from an agentic perspective,''
  \emph{Self-efficacy beliefs of adolescents}, vol.~5, pp. 1--43, 2006.

\bibitem{kegan1994}
R.~Kegan, \emph{In over our heads: The mental demands of modern life}.\hskip
  1em plus 0.5em minus 0.4em\relax Harvard University Press, 1994.

\bibitem{magolda2004making}
M.~B. Baxter~Magolda, \emph{Making their own way: Narratives for transforming
  higher education to promote self-development}.\hskip 1em plus 0.5em minus
  0.4em\relax Stylus Publishing, LLC., 2004.

\bibitem{willison2008researcher}
J.~Willison and K.~O’Regan, ``The researcher skill development framework,''
  \emph{https://www.adelaide.edu.au/melt/ua/media/51/rsd-framework.pdf}, 2008.

\bibitem{willison2012academics}
J.~W. Willison, ``When academics integrate research skill development in the
  curriculum,'' \emph{Higher Education Research \& Development}, vol.~31,
  no.~6, pp. 905--919, 2012.

\bibitem{willison2016phd}
J.~Willison and F.~Buisman-Pijlman, ``Phd prepared: research skill development
  across the undergraduate years,'' \emph{International Journal for Researcher
  Development}, vol.~7, no.~1, pp. 63--83, 2016.

\bibitem{Feldt:2009}
R.~Feldt, M.~Höst, and F.~Lüders, ``Generic skills in software engineering
  master thesis projects: Towards rubric-based evaluation,'' in \emph{2009 22nd
  Conference on Software Engineering Education and Training}, 2009, pp. 12--15.

\bibitem{kitchenham2002preliminary}
B.~A. Kitchenham, S.~L. Pfleeger, L.~M. Pickard, P.~W. Jones, D.~C. Hoaglin,
  K.~El~Emam, and J.~Rosenberg, ``Preliminary guidelines for empirical research
  in software engineering,'' \emph{IEEE Transactions on software engineering},
  vol.~28, no.~8, pp. 721--734, 2002.

\bibitem{Wohlin:12}
C.~Wohlin, P.~Runeson, M.~H{\"o}st, M.~C. Ohlsson, B.~Regnell, and
  A.~Wessl{\'e}n, \emph{Experimentation in software engineering}.\hskip 1em
  plus 0.5em minus 0.4em\relax Springer Science \& Business Media, 2012.

\bibitem{stol2020guidelines}
K.-J. Stol and B.~Fitzgerald, ``Guidelines for conducting software engineering
  research,'' in \emph{Contemporary Empirical Methods in Software
  Engineering}.\hskip 1em plus 0.5em minus 0.4em\relax Springer, 2020, pp.
  27--62.

\bibitem{bavota2012teaching}
G.~Bavota, A.~De~Lucia, F.~Fasano, R.~Oliveto, and C.~Zottoli, ``Teaching
  software engineering and software project management: An integrated and
  practical approach,'' in \emph{2012 34th international conference on software
  engineering (ICSE)}.\hskip 1em plus 0.5em minus 0.4em\relax IEEE, 2012, pp.
  1155--1164.

\bibitem{Hazzan:15}
O.~Hazzan, T.~Lapidot, and N.~Ragonis, \emph{Guide to teaching computer
  science: An activity-based approach}.\hskip 1em plus 0.5em minus 0.4em\relax
  Springer, 2015.

\bibitem{cain2013examining}
A.~Cain, C.~J. Woodward, and S.~Pace, ``Examining student progress in portfolio
  assessed introductory programming,'' in \emph{Proceedings of 2013 IEEE
  International Conference on Teaching, Assessment and Learning for Engineering
  (TALE)}.\hskip 1em plus 0.5em minus 0.4em\relax IEEE, 2013, pp. 67--72.

\bibitem{guest2022knowing}
J.~Guest and R.~Riegler, ``Knowing he standards: how good are students at
  evaluating academic work?'' \emph{Higher Education Research \& Development},
  vol.~41, no.~3, pp. 714--728, 2022.

\bibitem{Tubino:20}
L.~Tubino, A.~Cain, J.-G. Schneider, D.~Thiruvady, and N.~Fernando, ``Authentic
  individual assessment for team-based software engineering projects,'' in
  \emph{2020 IEEE/ACM 42nd International Conference on Software Engineering:
  Software Engineering Education and Training (ICSE-SEET)}.\hskip 1em plus
  0.5em minus 0.4em\relax IEEE, 2020, pp. 71--81.

\bibitem{yeates2013you}
P.~Yeates, P.~O'Neill, K.~Mann, and K.~W~Eva, ``‘you're certainly relatively
  competent’: assessor bias due to recent experiences,'' \emph{Medical
  education}, vol.~47, no.~9, pp. 910--922, 2013.

\bibitem{ernst2021understanding}
N.~A. Ernst, J.~C. Carver, D.~Mendez, and M.~Torchiano, ``Understanding peer
  review of software engineering papers,'' \emph{Empirical Software
  Engineering}, vol.~26, no.~5, pp. 1--29, 2021.

\bibitem{olorunnife2021automatic}
K.~Olorunnife, K.~Lee, and J.~Kua, ``Automatic failure recovery for
  container-based iot edge applications,'' \emph{Electronics}, vol.~10, no.~23,
  p. 3047, 2021.

\bibitem{wang2021virtual}
O.~Wang, B.~Cheng, T.~Hoang, C.~Arora, and X.~Liu, ``Virtual reality enabled
  human-centric requirements engineering,'' in \emph{2021 36th IEEE/ACM
  International Conference on Automated Software Engineering Workshops
  (ASEW)}.\hskip 1em plus 0.5em minus 0.4em\relax IEEE, 2021, pp. 159--164.

\bibitem{lefevre2022modelops}
K.~Lefevre, C.~Arora, K.~Lee, A.~Zaslavsky, M.~R. Bouadjenek, A.~Hassani, and
  I.~Razzak, ``Modelops for enhanced decision-making and governance in
  emergency control rooms,'' \emph{Environment Systems and Decisions}, pp.
  1--15, 2022.

\bibitem{CobbQuantum2022}
\BIBentryALTinterwordspacing
A.~Cobb, J.-G. Schneider, and K.~Lee, ``Towards higher-level abstractions for
  quantum computing,'' in \emph{Australasian Computer Science Week 2022}, ser.
  ACSW 2022.\hskip 1em plus 0.5em minus 0.4em\relax New York, NY, USA:
  Association for Computing Machinery, 2022, p. 115–124. [Online]. Available:
  \url{https://doi.org/10.1145/3511616.3513106}
\BIBentrySTDinterwordspacing

\bibitem{tubino2021reforming}
L.~Tubino, J.-G. Schneider, A.~Cain, D.~Thiruvady, and C.~Ranaweera,
  ``Reforming assessment: challenges beyond design,'' in \emph{2021 IEEE/ACM
  43rd International Conference on Software Engineering: Software Engineering
  Education and Training (ICSE-SEET)}.\hskip 1em plus 0.5em minus 0.4em\relax
  IEEE, 2021, pp. 78--88.

\bibitem{healey2009developing}
M.~Healey and A.~Jenkins, \emph{Developing undergraduate research and
  inquiry}.\hskip 1em plus 0.5em minus 0.4em\relax Higher Education Academy
  York, 2009.

\bibitem{matzen2012defining}
R.~Matzen and R.~Alrifai, ``Defining undergraduate research in computer
  science: a survey of computer science faculty,'' \emph{Journal of Computing
  Sciences in Colleges}, vol.~27, no.~3, pp. 31--37, 2012.

\bibitem{linn2015undergraduate}
M.~C. Linn, E.~Palmer, A.~Baranger, E.~Gerard, and E.~Stone, ``Undergraduate
  research experiences: Impacts and opportunities,'' \emph{Science}, vol. 347,
  no. 6222, p. 1261757, 2015.

\bibitem{mcmanus2019project}
J.~W. McManus and P.~J. Costello, ``Project based learning in computer science:
  a student and research advisor's perspective,'' \emph{Journal of Computing
  Sciences in Colleges}, vol.~34, no.~3, pp. 38--46, 2019.

\bibitem{paasivaara2018does}
M.~Paasivaara, D.~Vod{\u{a}}, V.~T. Heikkil{\"a}, J.~Vanhanen, and
  C.~Lassenius, ``How does participating in a capstone project with industrial
  customers affect student attitudes?'' in \emph{Proceedings of the 40th
  International Conference on Software Engineering: Software Engineering
  Education and Training}, 2018, pp. 49--57.

\bibitem{paasivaara2019collaborating}
M.~Paasivaara, J.~Vanhanen, and C.~Lassenius, ``Collaborating with industrial
  customers in a capstone project course: the customers' perspective,'' in
  \emph{2019 IEEE/ACM 41st International Conference on Software Engineering:
  Software Engineering Education and Training (ICSE-SEET)}.\hskip 1em plus
  0.5em minus 0.4em\relax IEEE, 2019, pp. 12--22.

\bibitem{bernat2000structuring}
A.~Bernat, P.~J. Teller, A.~Gates, and N.~Delgado, ``Structuring the student
  research experience,'' in \emph{Proceedings of the 5th annual SIGCSE/SIGCUE
  ITiCSEconference on Innovation and technology in computer science education},
  2000, pp. 17--20.

\bibitem{dodig2009professional}
G.~Dodig-Crnkovic and R.~Feldt, ``Professional and ethical issues of software
  engineering curricula,'' \emph{2010}, 2009.

\bibitem{dodig2010improved}
G.~Dodig-Crnkovic, F.~L{\"u}ders, M.~H{\"o}st, and R.~Feldt, ``Improved support
  for master’s thesis projects in software engineering,'' \emph{Rapport nr.:
  Rapporter fr{\aa}n NSHU}, 2010.

\bibitem{Host:2010}
M.~Höst, R.~Feldt, and F.~Luders, ``Support for different roles in software
  engineering master's thesis projects,'' \emph{IEEE Transactions on
  Education}, vol.~53, no.~2, pp. 288--296, 2010.

\bibitem{seyed2006practical}
B.~Seyed-Abbassi, ``Practical aspects of promoting research in a graduate
  course,'' \emph{Director}, p.~07, 2006.

\bibitem{shaw1998research}
M.~L. Shaw and B.~R. Gaines, ``A research-based masters program in the
  workplace,'' \emph{Proceedings of WCCCE}, vol.~98, p. 3rd, 1998.

\bibitem{bargh2014research}
M.~S. Bargh, A.~van Rooij-Peiman, L.~Remijn, and S.~Choenni, ``Research skills
  for software engineering undergraduates in dutch universities of applied
  sciences,'' in \emph{Proceedings of the 15th Annual Conference on Information
  technology education}, 2014, pp. 101--108.

\bibitem{khmelevsky2012automatic}
Y.~Khmelevsky, G.~Hains, and C.~Li, ``Automatic code generation within
  student's software engineering projects,'' in \emph{Proceedings of the
  Seventeenth Western Canadian Conference on Computing Education}, 2012, pp.
  29--33.

\bibitem{oyeniran2021environment}
A.~S. Oyeniran, T.~Ademilua, M.~Kruus, and R.~Ubar, ``Environment for
  innovative university research training in the field of digital test,'' in
  \emph{2021 30th Annual Conference of the European Association for Education
  in Electrical and Information Engineering (EAEEIE)}.\hskip 1em plus 0.5em
  minus 0.4em\relax IEEE, 2021, pp. 1--6.

\end{thebibliography}
